# Design of Synchronous Section-Carry Based Carry Lookahead Adders with Improved Figure of Merit

P. BALASUBRAMANIAN*
* School of Computer Science and Engineering
Nanyang Technological University
50 Nanyang Avenue
SINGAPORE 639798
balasubramanian@ntu.edu.sg

*Abstract:* - The section-carry based carry lookahead adder (SCBCLA) architecture was proposed as an efficient alternative to the conventional carry lookahead adder (CCLA) architecture for the physical implementation of computer arithmetic. In previous related works, self-timed SCBCLA architectures and synchronous SCBCLA architectures were realized using standard cells and FPGAs. In this work, we deal with improved realizations of synchronous SCBCLA architectures designed in a semi-custom fashion using standard cells. The improvement is quantified in terms of a figure of merit (FOM), where the FOM is defined as the inverse product of power, delay and area. Since power, delay and area of digital designs are desirable to be minimized, the FOM is desirable to be maximized. Starting from an efficient conventional carry lookahead generator, we show how an optimized section-carry based carry lookahead generator is realized. In comparison with our recent work dealing with standard cells based implementation of SCBCLAs to perform 32-bit addition of two binary operands, we show in this work that with improved section-carry based carry lookahead generators, the resulting SCBCLAs exhibit significant improvements in FOM. Compared to the earlier optimized hybrid SCBCLA, the proposed optimized hybrid SCBCLA improves the FOM by 88.3%. Even the optimized hybrid CCLA features improvement in FOM by 77.3% over the earlier optimized hybrid CCLA. However, the proposed optimized hybrid SCBCLA is still the winner and has a better FOM than the currently optimized hybrid CCLA by 15.3%. All the CCLAs and SCBCLAs are implemented to realize 32-bit dual-operand binary addition using a 32/28nm CMOS process.

## 1 Introduction
The carry lookahead adder (CLA) is an important member of the high-speed digital adder family [1] [2] and is a logarithmic time adder. In the existing literature, the CLA has been implemented in different logic styles such as static CMOS [3], dynamic CMOS [4] [5], all N-transistor logic [6], pass transistor logic [7], adiabatic style energy recovery [8] [9], gate-diffusion input [10] [11], quantum dot cellular automata (QCA) [12], and using a variety of materials such as gallium arsenide [13], memristor [14] and vertically-stacked nanowire transistors [15] besides the standard Si-based CMOS. Further, different design styles were considered for the CLA implementation such as self-timed [16 – 18] and synchronous viz. full-custom and semi-custom ASIC and FPGA based designs [19 – 22].

The design of a CLA is based on the principle that by examining the augend and addend inputs of an adder, it is possible to predict the carry signal of any arbitrary adder stage a priori thus paving the way for significant reduction of linear propagation delay that would be manifest in a ripple carry adder (RCA) architecture, where the carry signal tends to serially propagate from one full adder stage to the next. There are two types of CLA architectures: homogeneous and hybrid/heterogeneous. Pure CLAs are called homogeneous CLAs while hybrid CLAs contain a combination of CLAs and other carry-propagate adder architectures, for example CLA and RCA.

The remaining part of this article is organized as follows. Section 2 discusses the CCLA architecture. Section 3 describes the SCBCLA architecture. Sections 2 and 3 also highlight the physical difference between the carry lookahead generators considered in this work and in the previous work [19]. It has been shown in our prior work [19] that hybrid CCLAs and SCBCLAs fare well compared to their homogeneous counterparts with respect to optimization of design metrics viz. power, delay and area. Section 4 presents the simulation results corresponding to different homogeneous and hybrid CCLAs and SCBCLAs, which are suitable for performing 32-bit dual-operand binary addition. Finally, the conclusions are given in Section 5.

## 2 Conventional Carry Lookahead Generator and Adder

Let us assume that $A_i$ and $B_i$ are the augend and addend inputs of an adder i.e. full adder stage, and $C_i$ is its carry input. The (lookahead) carry output viz. $C_{i+1}$ is then expressed by (1), and the sum output is expressed by (2).

$$C_{i+1} = G_i + P_iC_i \tag{1}$$

$$Sum_i = P_i \oplus C_i \tag{2}$$

The binary full adder [23 – 25] is a fundamental arithmetic unit that adds two input bits inclusive of any incoming carry input and produces the sum and carry (overflow) output. In (1) and (2), $G_i$ and $P_i$ represent generate and propagate signals, where $G_i = A_iB_i$ and $P_i = A_i \oplus B_i$. Product implies logical conjunction, and sum implies logical disjunction in the equations. The symbol $\oplus$ specifies logical exclusivity (i.e. logical XOR). Notice that generate and propagate functions are mutually exclusive – hence the carry is either generated from an adder stage or the carry simply propagates from the input to output. Equations (1) and (2) are inherently in disjoint sum of products (DSOP) or sum of disjoint products form [26] [27]. In such a form, any two product terms constituting the Boolean expression would be mutually orthogonal [28] [29], i.e. the logical conjunction of any pair-wise combination of the product terms would equate to null (i.e. binary 0).

Unwinding the recursion implicit in (1), the carry lookahead outputs corresponding to a 4-bit carry lookahead generator are specified by (3) to (6), where $C_0$ represents the carry input to the 4-bit carry lookahead generator, and $C_1$, $C_2$, $C_3$ and $C_4$ represent the corresponding lookahead carry outputs generated. Notice that in a generic m-bit CCLA, a total of 'm' lookahead carry outputs are produced. Equations (3), (4), (5) and (6) show how the lookahead carries are dependent only upon the incoming carry-input to the carry lookahead generator and the corresponding generate and propagate signals, i.e. there is no relation between the intermediate carries. Equations (3), (4), (5) and (6) are inherently in DSOP form.

$$C_4 = G_3 + P_3G_2 + P_3P_2G_1 + P_3P_2P_1G_0 + P_3P_2P_1P_0C_0 \tag{3}$$

$$C_3 = G_2 + P_2G_1 + P_2P_1G_0 + P_2P_1P_0C_0 \tag{4}$$

$$C_2 = G_1 + P_1G_0 + P_1P_0C_0 \tag{5}$$

$$C_1 = G_0 + P_0C_0 \tag{6}$$

Fig. 1 portrays the architecture of a generic m-bit CCLA. It consists of 3 parts: (i) propagate-generate logic, which produces propagate and generate signals corresponding to the augend and addend inputs, (ii) a m-bit conventional carry lookahead generator, which accepts the propagate and generate signals and the carry input ($C_0$) and processes them to produce the lookahead carry outputs including the carry input for the successive CLA, and (iii) the sum logic, which combines the respective propagate and carry signals of the m-bit CCLA according to (2) and processes them to produce the sum outputs of the CCLA.

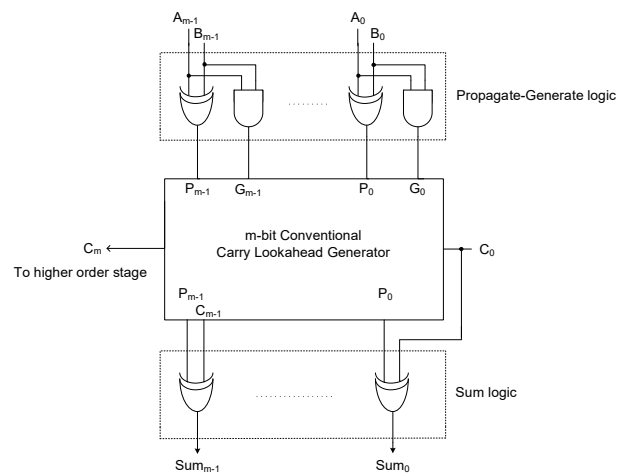

Fig. 1 Microarchitecture of m-bit CCLA

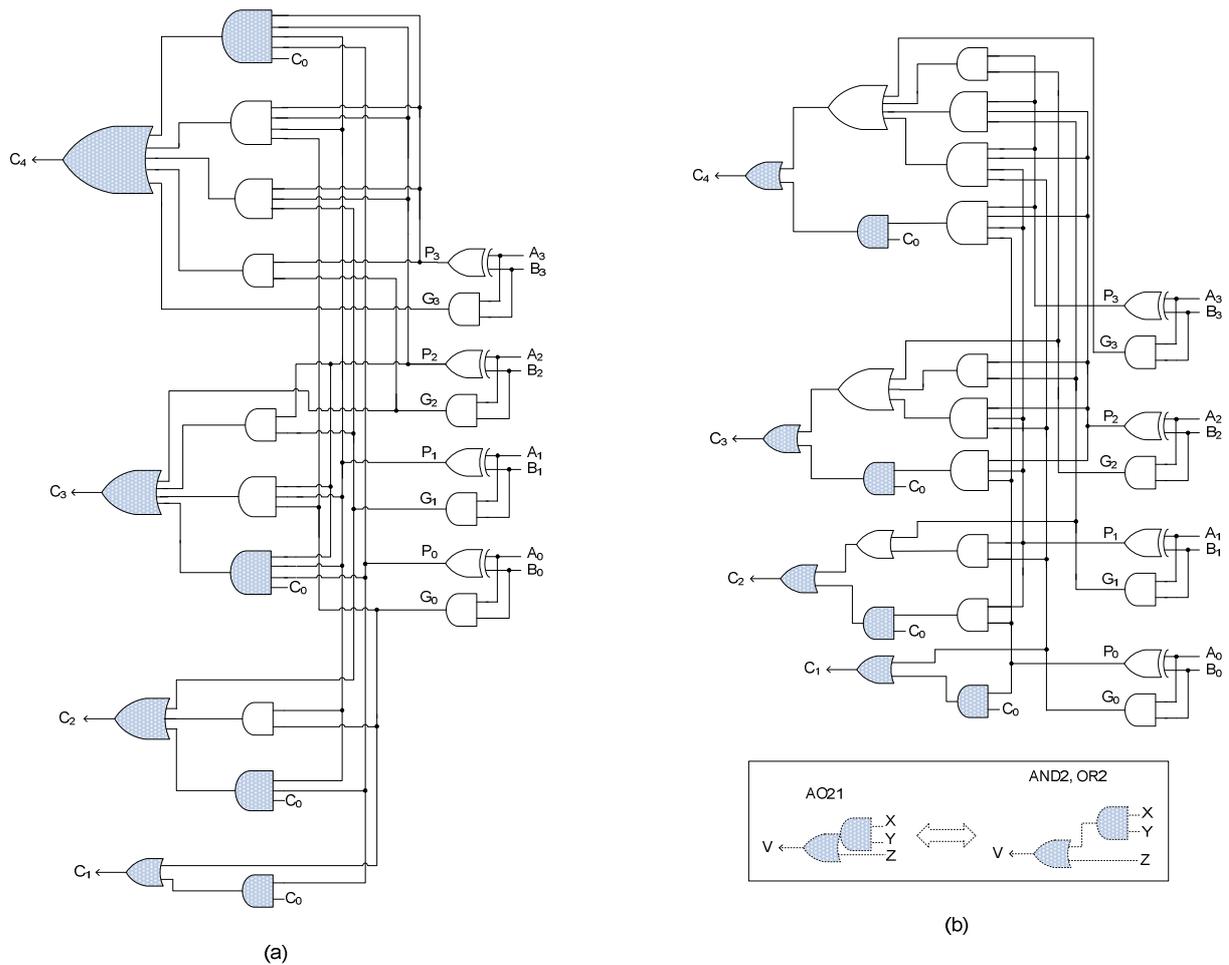

Fig. 2 Example 4-bit conventional carry lookahead generators
(a) Basic implementation, (b) Decomposed (i.e. carry optimized) implementation

Fig. 2 shows two structural implementations of an example 4-bit conventional carry lookahead generator viz. basic implementation shown as Fig. 2a, and the decomposed implementation shown as Fig. 2b. Both these implementations synthesize (3) to (6). An example decomposed carry lookahead generator given in [30] has been considered here and is further optimized using complex gates in this work. The decomposed implementation can be obtained through logic factoring [31 – 33] of the basic implementation which would help to optimize the design parameters – a generalized illustration is given in the Appendix.

The basic 4-bit conventional carry lookahead generator shown in Fig. 2a cannot be physically realized as such using modern CMOS technologies which are less than 90nm since the fan-in of the AND/NAND gates and OR/NOR gates are usually restricted to maximum of 4. It can be seen in Fig. 2a that $C_4$ requires a final 5-input OR gate, which is not available in a modern cell library [34]. Hence, a better alternative is to opt for a decomposed conventional carry lookahead generator which can not only be implemented using a modern digital cell library but can also pave the way for minimizing the propagation delays in generating the lookahead carry outputs. For example, in the case of the basic 4-bit conventional carry lookahead generator shown in Fig. 2a, and assuming that this carry lookahead generator based sub-CCLA is used in an intermediate stage of an n-bit CCLA, it can be seen that the propagation delays associated with the lookahead carry outputs would tend to grow proportionate to the gates sizes. This is visually evident from the shaded gates shown in Fig. 2a. For the basic implementation of a conventional carry lookahead generator, an example of which is shown in Fig. 2a, the propagation delay incurred in generating a $K^{th}$ lookahead carry output viz. $C_K$ would equal the sum of propagation delays of a (K+1)-input AND gate and a (K+1)-input OR gate whilst presuming that the basic conventional carry lookahead generator is present in some arbitrary internal stage of an n-bit CCLA.

On the other hand, assuming that the decomposed 4-bit conventional carry lookahead generator shown in Fig. 2b is present in some intermediate sub-CCLA of an n-bit CCLA, it can be seen from Fig. 2b that the propagation delays incurred in generating all the lookahead carry outputs are consistent and are a minimum, which equate to the sum of propagation delays of a 2-input AND gate and a 2-input OR gate, which are shown shaded in Fig. 2b. The rectangular box at the bottom of Fig. 2b shows how a 2-input AND gate and a 2-input OR gate functionality can be merged into a single AO21 complex gate. Recall that both simple and complex gates are made available as part of digital standard cell libraries. By adopting this logic optimization at the technology mapping stage, the delays involved in generating the lookahead carry outputs of the logic decomposed conventional carry lookahead generator would be further minimized and would be equal to the propagation delay of a single AO21 gate. Thus the decomposed conventional carry lookahead generator implementation in Fig. 2b is highly favorable for delay optimization compared to the basic implementation shown in Fig. 2a. Also, the decomposed implementation makes it possible to physically realize a conventional carry lookahead generator of any size whilst paving the way for a consistent and optimized delay in generating all the lookahead carry outputs, which is not the case with the basic implementation.

Fig. 3 shows 5 different homogeneous and hybrid CCLA architectures suitable for performing 32-bit dual-operand binary addition. The introduction of a RCA in the least significant position in a hybrid CCLA could help to reduce its critical path delay since the most significant lookahead carry output of a 4-bit CCLA present in the least significant position of a hybrid n-bit CCLA would encounter a 2-input XOR gate, a 4-input AND gate, a 4-input OR gate and a 2-input OR gate in its critical path as seen in Fig. 2b. The delay encountered in producing the most significant lookahead carry output from a sub 4-bit CCLA present in the least significant position could be compensated by positioning a small size RCA in the least significant position of a hybrid CCLA. This may not only help to reduce the delay of a hybrid CCLA but also could minimize its area and power. However, any overuse of RCA in the least significant position(s) might exacerbate the hybrid CCLA delay.

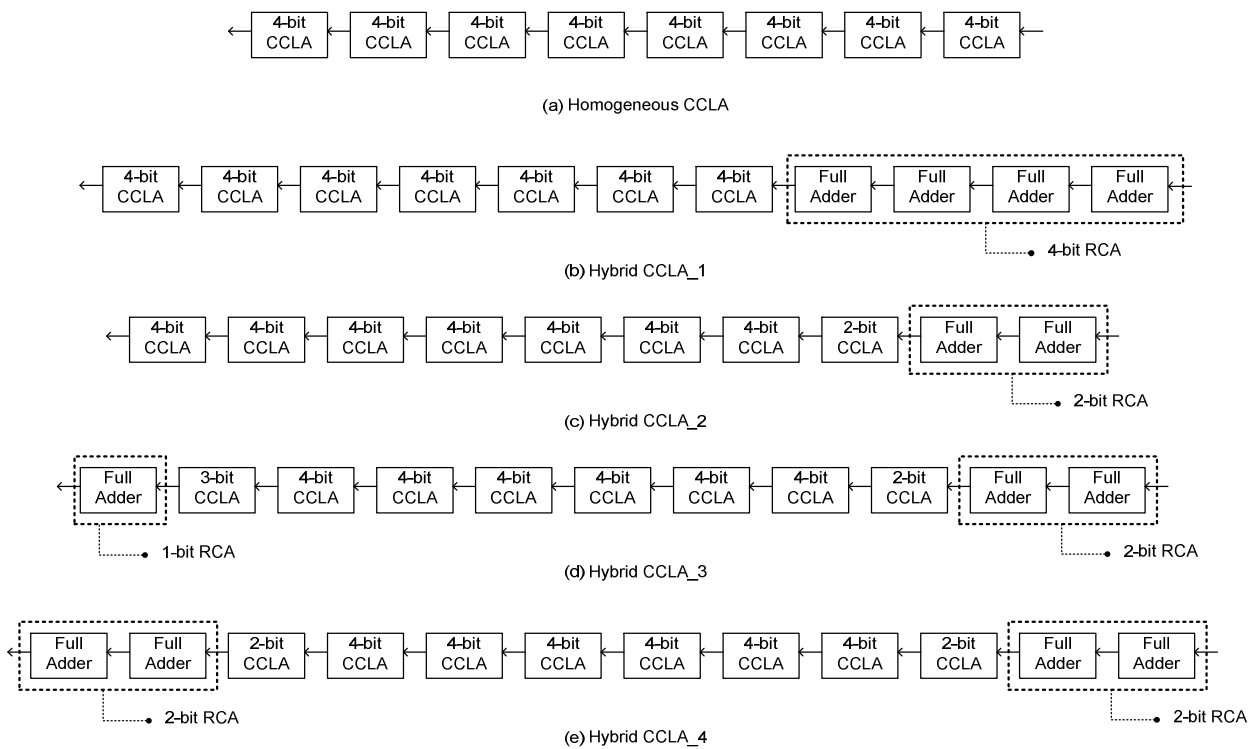

Fig. 3 Different homogeneous and hybrid 32-bit CCLA realizations. 3-bit CCLA blocks comprising a basic 3-bit conventional carry lookahead generator was widely used in [19]. In contrast, 4-bit CCLA blocks incorporating a decomposed 4-bit conventional carry lookahead generator is widely used here.

Fig. 3a shows a 32-bit homogeneous CCLA constructed using 8 numbers of 4-bit CCLAs. Fig. 3b shows a 32-bit hybrid CCLA comprising 7 numbers of 4-bit CCLAs and a 4-bit RCA in the least significant nibble position, labelled as Hybrid CCLA_1. Fig. 3c shows Hybrid CCLA_2, which comprises 7 numbers of 4-bit CCLAs, a 2-bit CCLA and a final 2-bit RCA in the least significant nibble position. Fig. 3d shows Hybrid CCLA_3, constructed using a 1-bit RCA and a 3-bit CCLA in the most significant nibble position, followed by 6 numbers of 4-bit CCLAs, which are then followed by a 2-bit CCLA and a 2-bit RCA in the least significant nibble position. Finally, the Hybrid CCLA_4 shown in Fig. 3e consists of a 2-bit RCA and a 2-bit CCLA in the most significant nibble position, followed by 6 numbers of 4-bit CCLAs, which are subsequently followed by a 2-bit CCLA and a 2-bit RCA in the least significant nibble position.

## 3 Section-Carry Based Carry Lookahead Generator and Adder

The SCBCLA is a derivative of the CCLA in that only one lookahead carry output, which could serve as the carry input for the successive SCBCLA stage is alone produced. The production of sum outputs of the SCBCLA is not dependent upon the internal lookahead carry outputs, instead they are produced based on the internal rippling of the carry input signal from one full adder stage to the next within the SCBCLA (i.e. sub-SCBCLA) as in the RCA. The SCBCLA receives the augend, addend and carry inputs and processes them to produce the corresponding sum outputs and a single lookahead carry output. The SCBCLA generates a lookahead carry output on the basis of the carry lookahead adder architecture while producing the sum outputs on the basis of the RCA architecture. The generic SCBCLA architecture is portrayed by Fig. 4.

As is the case with the generic CCLA architecture, the SCBCLA architecture also comprises 3 parts viz. the propagate-generate logic, the m-bit section-carry based carry lookahead generator, and the sum producing logic as shown in Fig. 4. Further, as mentioned earlier for the case of CCLAs, pure SCBCLAs are called homogeneous SCBCLAs, and hybrid/heterogeneous SCBCLAs feature SCBCLA and another adder architecture (for example, RCA). Moreover, heterogeneous SCBCLAs are indeed likely to feature optimized design metrics compared to homogeneous SCBCLAs. Topologically, the propagate-generate logic of the CCLA and SCBCLA is similar, but the carry lookahead generator and the sum producing logic are different as can be observed by comparing Fig. 1 with Fig. 4 shown below.

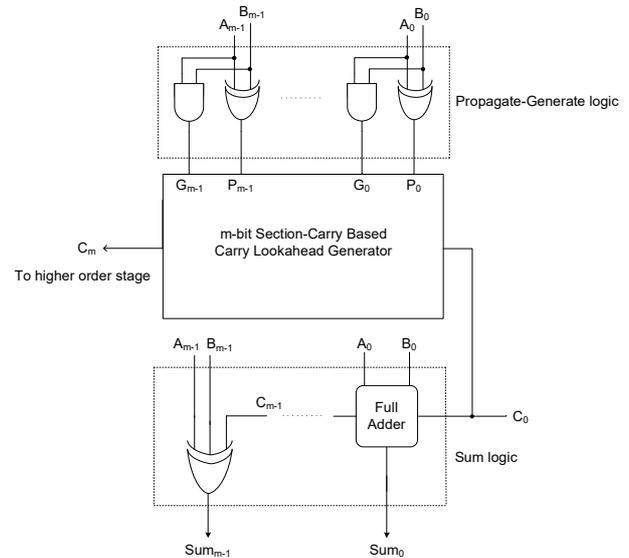

Fig. 4 Microarchitecture of m-bit SCBCLA

Fig. 5 shows the respective basic and decomposed implementations of an example 4-bit section-carry based carry lookahead generator whose carry output synthesizes (3). For the basic implementation, the propagation delay incurred in producing the lookahead carry output would be equal to the sum of propagation delays of a 2-input XOR gate, a 5-input AND gate and a 5-input OR gate as seen in Fig. 5a. For the decomposition implementation of Fig. 5b, the maximum delay encountered would equal the sum of propagation delays of a 2-input XOR gate, a 4-input AND gate, a 4-input OR gate and a 2-input OR gate.

It can be noticed that the basic implementation of the 4-bit section-carry based carry lookahead generator shown in Fig. 5a comprises 5-input AND and OR gates, which are not supported in modern CMOS process technologies [34]. This was found to be the problem with the direct implementation of a 4-bit conventional carry lookahead generator earlier in Fig. 2a. Therefore, a better alternative is to opt for the decomposed 4-bit section-carry based carry lookahead generator portrayed through Fig. 5b. It can be seen in Fig. 5a that upon receipt of the carry input viz. $C_0$, the production of the lookahead carry output will encounter the delay of a 5-input AND gate and a 5-input OR gate which are shown shaded. On the other hand, with reference to Fig. 5b, upon receipt of the carry input, the production of the lookahead carry output will encounter the delay of just a 2-input AND gate and a 2-input OR gate, whose delay is much less than the former. The recurring 2-input AND and OR gates in the critical path are shown shaded in Fig. 5b.

WSEAS TRANSACTIONS on CIRCUITS and SYSTEMS     P. BalasubramanianWSEAS TRANSACTIONS on CIRCUITS and SYSTEMS     P. Balasubramanian

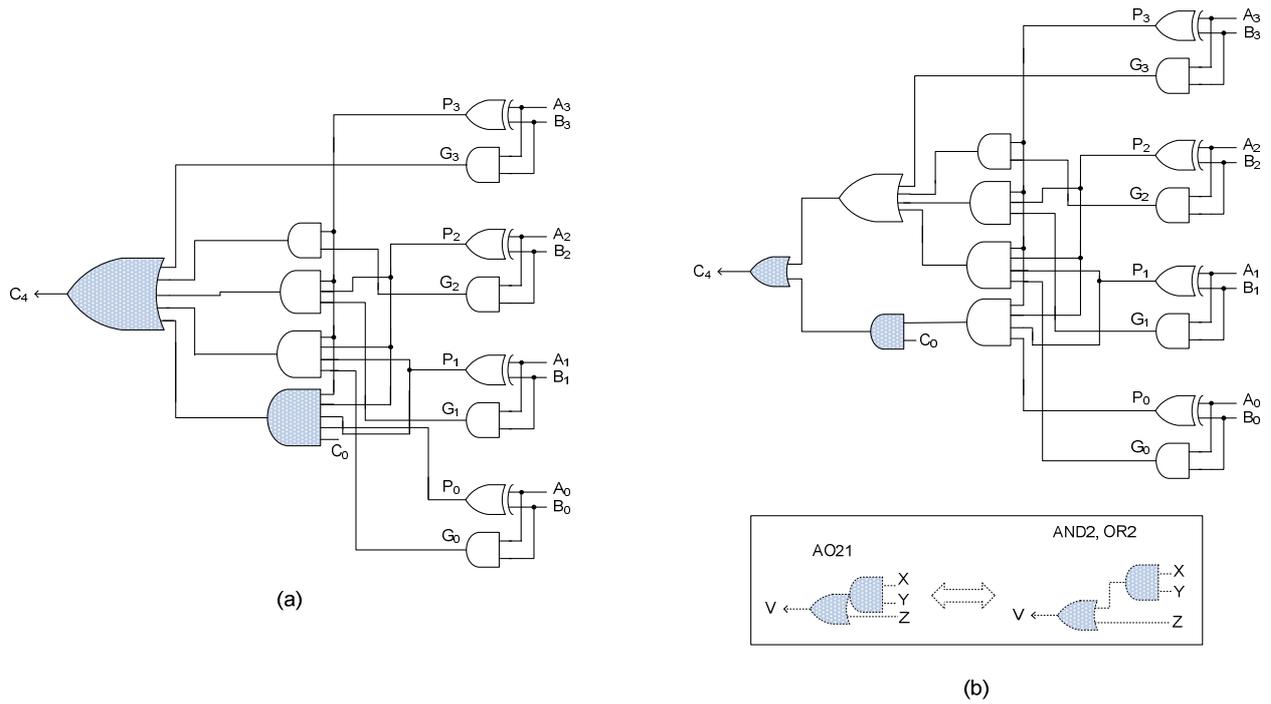

Fig. 5 Example 4-bit section-carry based carry lookahead generators
(a) Basic implementation, (b) Decomposed (i.e. carry optimized) implementation

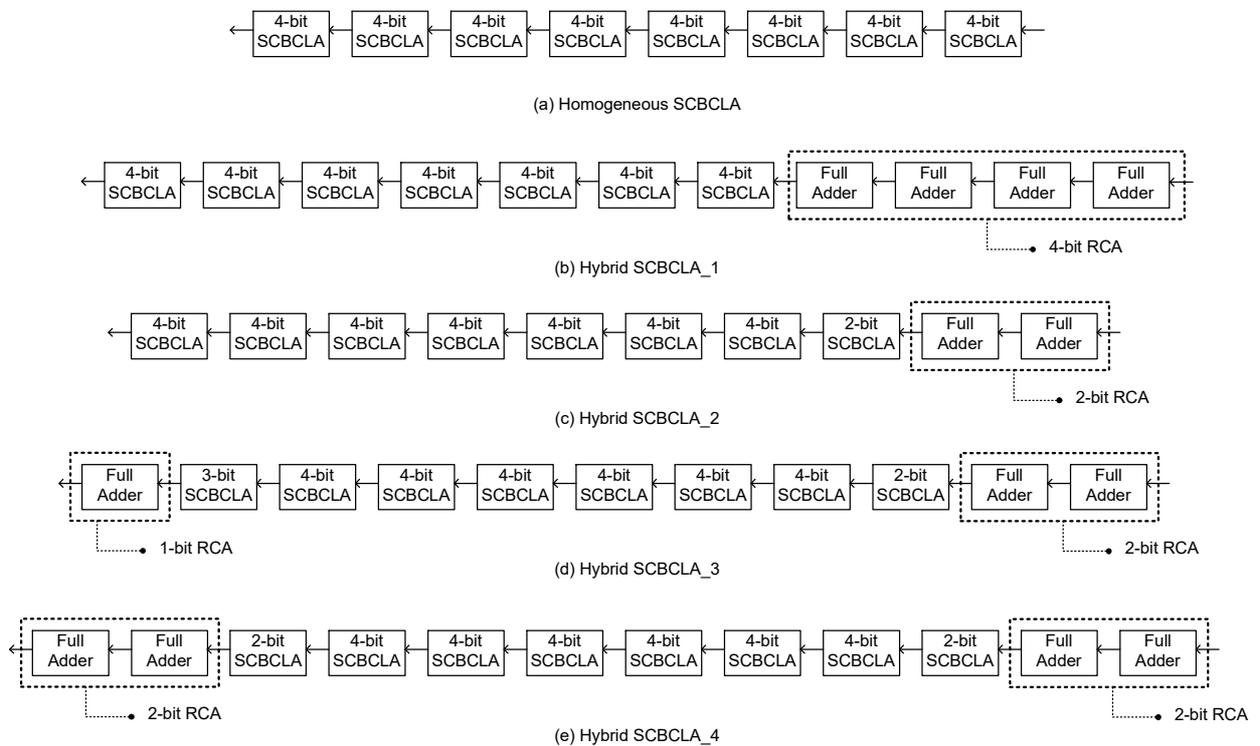

Fig. 6 Various homogeneous and hybrid 32-bit SCBCLA realizations. 3-bit SCBCLA blocks comprising a basic 3-bit section-carry based carry lookahead generator was widely used in [19]. In contrast, 4-bit SCBCLA blocks comprising a decomposed 4-bit section-carry based carry lookahead generator is widely used in this work. Note the architectures of homogeneous and hybrid CCLAs and SCBCLAs are similar.

Moreover, the carry output logic can be optimized by merging the 2-input AND gate and the 2-input OR gate into a single AO21 complex gate as portrayed by the rectangular box appearing at the bottom of Fig. 5b. Thus the decomposed implementation of the section-carry based carry lookahead generator shown in Fig. 5b is highly favorable for delay optimization compared to the basic implementation shown in Fig. 5a. Further, the decomposed implementation makes it possible to physically realize a section-carry based carry lookahead generator of any size whilst paving the way for a consistent and optimized delay in generating the lookahead carry output, which is not the case with the basic implementation.

Fig. 6 shows the topologies of different 32-bit homogeneous and hybrid SCBCLAs. These are identical to the homogeneous and hybrid CCLA topologies shown in Fig. 3. Fig. 6a shows a 32-bit homogeneous SCBCLA topology, while Figs. 6b to 6e depict the topologies of Hybrid SCBCLA_1 to Hybrid SCBCLA_4 respectively.

## 4 Results and Inferences

Standard cell based semi-custom ASIC style implementation of different homogeneous and heterogeneous 32-bit CCLAs and SCBCLAs was considered in this work for comparison purpose. The CCLAs and SCBCLAs were realized using a high $V_t$ 32/28nm CMOS process [34]. The power, delay and area estimates corresponding to 32-bit CCLAs are given in Table 1, and the design metrics estimated for 32-bit SCBCLAs are given in Table 2.

The design metrics estimated correspond to a typical case PVT specification with recommended supply voltage of 1.05V and operating junction temperature of 25ºC. For estimating the average power dissipation, more than 1000 random input vectors were identically supplied to the different CLAs at time intervals of 5ns (200MHz) through a test bench, similar to that of [19]. The value change dump (.vcd) files generated through the functional simulations were subsequently utilized to accurately estimate the average power dissipation through Synopsys PrimeTime by invoking the time-based power analysis mode. The maximum propagation delay (i.e. critical path delay) and area occupancy were also estimated with suitable wire loads included automatically whilst performing the simulations. Minimum-sized discrete and complex gates of the digital cell library [34] were chosen uniformly for realizing the different CCLAs and SCBCLAs. This in fact paves the way for a direct comparison of the design metrics of different CCLAs and SCBCLAs post-physical synthesis.

Table 1. Average power dissipation, critical path delay, and Silicon area of different 32-bit CCLAs

| Type of CCLA | Power (µW) | Delay (ns) | Area (µm²) |
|---|---|---|---|
| Homogeneous CCLA (Fig. 3a) | 40.77 | 1.13 | 646.54 |
| Hybrid CCLA_1 (Fig. 3b) | 39.24 | 1.18 | 585.04 |
| Hybrid CCLA_2 (Fig. 3c) | 39.66 | 1.05 | 607.91 |
| Hybrid CCLA_3 (Fig. 3d) | 39.13 | 1.08 | 586.56 |
| Hybrid CCLA_4 (Fig. 3e) | 38.63 | 1.18 | 569.28 |

Table 2. Average power dissipation, critical path delay, and Silicon area of different 32-bit SCBCLAs

| Type of SCBCLA | Power (µW) | Delay (ns) | Area (µm²) |
|---|---|---|---|
| Homogeneous SCBCLA (Fig. 6a) | 43.64 | 1.26 | 500.16 |
| Hybrid SCBCLA_1 (Fig. 6b) | 41.82 | 1.32 | 456.95 |
| Hybrid SCBCLA_2 (Fig. 6c) | 42.44 | 1.19 | 480.59 |
| Hybrid SCBCLA_3 (Fig. 6d) | 42.08 | 1.11 | 470.17 |
| Hybrid SCBCLA_4 (Fig. 6e) | 41.63 | 1.12 | 461.02 |

Firstly, it is noted that the least critical path delay of the 32-bit hybrid CCLA in the previous work [19] is 2.18ns, whereas Table 1 shows that the minimal critical path delay of Hybrid CCLA_2 is just 1.05ns, which amounts to a 51.8% reduction in delay or in other words, a 107.6% improvement in speed for the latter compared to the former, albeit at the expense of relatively moderate increases in area and power. This significant improvement in speed for the hybrid CCLA of this work compared to the hybrid CCLA of the previous work is facilitated due to 2 important reasons: i) choosing a decomposed implementation for realizing a conventional carry lookahead generator in this work as opposed to the choice of a basic/direct implementation for realizing a conventional carry lookahead generator in the previous work, and ii) the wide usage of a sub 4-bit CCLA to construct a 32-bit hybrid CCLA in this work instead of the extensive use of a sub 3-bit CCLA to construct a 32-bit hybrid CCLA as done in the previous work.

Secondly, a similar observation is made with respect to the hybrid SCBCLAs of this work in comparison with those of the previous work. It is noted that the least critical path delay of the 32-bit hybrid SCBCLA in the previous work is 2.16ns. In contrast, the least critical path delay of Hybrid SCBCLA_3 is just 1.11ns, which amounts to a 48.6% reduction in delay for the latter or in other words, a 94.6% improvement in speed for the latter compared to the former. Nevertheless, this reduction in critical path delay or increase in speed is achieved at the expense of just moderate increases in area and power.

To holistically comment on the design parameters viz. power, delay and area of different CCLAs and SCBCLAs, a figure-of-merit (FOM) is defined as the inverse product of power, delay, and area as in [19], [35 – 41]. Since minimization of power, delay, and area is desirable, a lower power-delay-area product and thus a higher FOM is an indicator of an optimized design. The calculated FOM values of various 32-bit CCLAs and SCBCLAs, which are scaled up by a factor of $10^6$, are portrayed through Figs. 7 and 8 respectively.

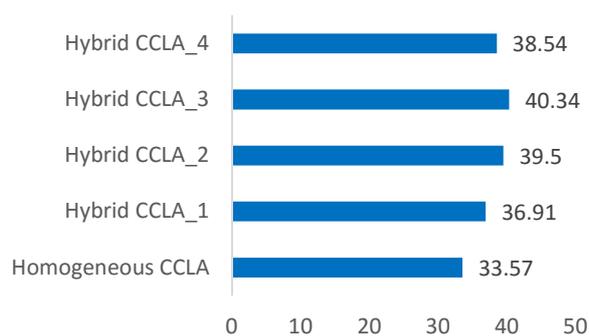

Fig. 7 FOM of different 32-bit CCLAs

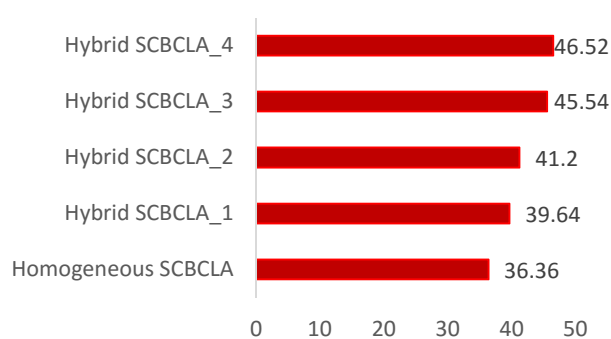

Fig. 8 FOM of different 32-bit SCBCLAs

The best FOM achieved by the hybrid CCLA in the previous work is just 22.5. In contrast, the best FOM achieved by Hybrid CCLA_3 (of this work) is 40.34 as seen in Fig. 7, which signifies a 79.3% increase. On similar lines, the best FOM achieved by the hybrid SCBCLA in the previous work is just 24.7, while the best FOM achieved by Hybrid SCBCLA_4 (in this work) is 46.52 as seen in Fig. 8 which signifies an 88.3% increase. Even with respect to homogeneous CCLAs and SCBCLAs, this work reports corresponding increases in FOM compared to the previous work. For example, the homogeneous 32-bit CCLA of this work reports approximately 60% increase in FOM compared to the homogeneous 32-bit CCLA of the previous work. Overall, from Figs. 7 and 8, it can be inferred that the Hybrid SCBCLA_4 reports the best FOM of all the CLAs implemented, and has an edge over the FOM of Hybrid CCLA_3 by 15.3%.

The final conclusions are: i) logic decomposed implementations of carry lookahead generators of any kind (referring to Figs. 2b and 5b) are indeed practically viable and physically advantageous over basic implementations of carry lookahead generators (referring to Figs. 2a and 5a), ii) hybrid CLAs of any kind (i.e. CCLAs or SCBCLAs) generally exhibit superior FOM than their counterpart homogeneous CLAs, iii) from a FOM perspective, primarily, hybrid SCBCLAs are preferable compared to homogeneous SCBCLAs as well as homogeneous/hybrid CCLAs, and iv) from an individual design metric perspective, hybrid CCLAs or SCBCLAs may be preferable, and this has to be decided based on the actual design requirement(s).

APPENDIX

The following is a generalized illustration of how to synthesize a decomposed carry lookahead output equation with minimum stage delay.

Unwinding the recursion implicit in (1), we obtain

$$C_{i+1} = G_i + P_i G_{i-1} + P_i P_{i-1} G_{i-2} + \ldots + P_i P_{i-1} P_{i-2} \bullet \ldots \bullet C_0$$

Where, 'i' represents any intermediate stage within a CCLA/SCBCLA, $C_{i+1}$ denotes the carry lookahead output of the corresponding $i^{th}$ stage, and $C_0$ represents the carry input to the CCLA/SCBCLA. Product implies logical conjunction and sum implies logical disjunction. The symbol '+' signifies Boolean sum, and the symbol '$\bullet$' signifies Boolean product.

Let $K = P_i G_{i-1}$, $L = P_i P_{i-1} G_{i-2}$, and $M = P_i P_{i-1} P_{i-2} \bullet \ldots$

Therefore, $C_{i+1} = G_i + K + L + \ldots + MC_0$

Further, let $N = G_i + K + L + \ldots$

Hence we have, $C_{i+1} = N + MC_0$

The propagate (P) and generate (G) signals are realized using XOR and AND gates respectively. The intermediate variables K, L and M are realized using non-decomposed/decomposed AND gates and the intermediate variable N is realized using non-decomposed/decomposed OR gate(s), subject to the cell library constraints [34].

The final carry lookahead output ($C_{i+1}$) equation given above involving N, M and $C_0$ is synthesized using a single complex gate viz. the AO21 as shown below. In the below realization, the carry input is subject to a minimal critical path involving just a single complex gate.

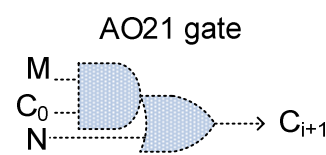